\documentclass[11pt]{article}
\usepackage{amsmath}
\usepackage{graphicx}
\usepackage{amsfonts}
\usepackage{amssymb}
\usepackage{epsfig}
\usepackage{color}
\usepackage{psfrag}
\usepackage{epstopdf}
\usepackage{comment}
\usepackage{caption}
\usepackage{subcaption}
\usepackage{braket}
\usepackage{multirow}
\usepackage{bbm}
\usepackage{algpseudocode}
\usepackage{algorithm}
\usepackage{url}
\usepackage{hyperref}

\newcommand{\EQ}{\begin{equation}}
\newcommand{\EN}{\end{equation}}
\newcommand{\be}{\begin{equation}}
\newcommand{\ee}{\end{equation}}
\newcommand{\bea}{\begin{eqnarray}}
\newcommand{\eea}{\end{eqnarray}}

\begin{document} \setcounter{page}{0}
\topmargin 0pt
\oddsidemargin 5mm
\renewcommand{\thefootnote}{\arabic{footnote}}
\newpage
\setcounter{page}{0}
\topmargin 0pt
\oddsidemargin 5mm
\renewcommand{\thefootnote}{\arabic{footnote}}
\newpage

\begin{center}
{\large {\bf Modeling Perpetrators' Fate-to-Fate Contagion in Public Mass Shootings In The United States Using  Bivariate Hawkes Processes}}\\
\vspace{1.8cm}
{\large Youness Diouane $^1$, James Silver $^2$}\\ \vspace{1cm}
{$^1$ \textit{Boston College, Computer Science Department, 245 Beacon St, Chestnut Hill, MA 02467, USA}}\\
{$^2$\textit{Boston University Metropolitan College, Criminal Justice Department, 1010 Commonwealth Ave, Boston, MA 02215, USA}}

\end{center}
\vspace{1.2cm}

\renewcommand{\thefootnote}{\arabic{footnote}}
\setcounter{footnote}{0}

\begin{abstract}

This study examines how the fate of a perpetrator in a public mass shooting influences the fate of subsequent perpetrators. Using data from 1966 to 2024, we classify incidents according to whether the perpetrator died at the scene or survived the attack. Using a bivariate Hawkes process, we quantify the cross-excitation effect, which is the triggering effect that each event type exerts on the other, i.e., "die at the scene"$\rightarrow$ "live" and "live"$\rightarrow$ "die at the scene", as well as the self-excitation effects, i.e., "die at the scene"$\rightarrow$ "die at the scene" and "live"$\rightarrow$ "live". Our results show that the strongest spillover was from "live" incidents to "die at the scene", where we estimate that 0.34 (0.09, 0.80) of "die at the scene" incidents are triggered by a prior event in which the offender survived the attack. This pathway also exhibits the longest estimated contagion timescale: approximately 20 days. In contrast, the reverse influence, that is, "die at the scene"$\rightarrow$"live", is not statistically significant, with the lower bound of its 95\% confidence interval nearly equal to zero. We also find that "die at the scene" events can only cause their own type, where 0.139 (0.01, 0.52) of such incidents are caused by previous "die at the scene" events, with the shortest contagion timescale of roughly 20 hours. When we analyze the pre and post 2000 periods separately, we observe a clear shift in dynamics. Before the year 2000, cross-excitation effects were approximately symmetric, with about 0.1 (0.003, 0.3) of incidents of either type triggered by a single event of the other type, and contagion timescales of roughly 2.5 days in both directions. This symmetry disappears after 2000. In the post 2000 era, the strongest spillover runs from "live" to "die at the scene", estimated at 0.228 (0.03, 0.55), with the longest contagion duration of about 4.3 days. The reverse effect is not statistically significant. Instead, as in the full-period analysis, "die at the scene" events trigger only their own type, with an estimated effect of 0.044 (0.01, 0.09) and the shortest contagion timescale of about 16 hours. This mirrors the pattern found in the full 1966--2024 dataset, which suggests that the dynamics of public mass shootings in the United States are mainly driven by the post-2000 era.

\end{abstract}
\section*{Introduction}

While public mass shootings are rare, their social impact is profound, prompting urgent calls for prevention. Amid ongoing research, a critical question has emerged: can the occurrence of one event influence the likelihood or characteristics of others, and if so, how? Exploring this question requires considering how reactions to these attacks - media coverage and public discourse – might contribute to patterns that are not immediately obvious. Our research seeks to illuminate whether and how the fate of offenders at the scene may play a role in shaping subsequent events, offering new insights into the dynamics of these signal crimes.

Defined here as incidents in which four or more people (not including the offender) are shot and killed within 24 hours in a public setting and which are not attributable to underlying criminal activity or commonplace circumstances \cite{krouse2015mass}, public mass shootings generate intense public concern and legislative interest \cite{chermak1998predicting, duwe2000body, duwe2004patterns, fox2021newsworthiness, fridel2022integrating, schildkraut2016mass}. Scholarly efforts to understand the phenomenon have ranged from creating offender typologies to researching specific antecedent behaviors and experiences (e.g., mental illness, masculinity norms), as well as life-course perspectives, all of which underpin the growing consensus that offenders do not just “snap,” but instead move over time along a “pathway to violence” \cite{silver2022sequence}. 

Based in part on the enormous media and public attention given to these events, researchers have also begun to examine the possibility of a contagion effect \cite{fox2021contagion}. Evidence of contagion has been found across various unrelated behaviors such as airplane hijacking \cite{holden1986contagiousness}, smoking cessation \cite{christakis2008collective}, and especially suicide \cite{abrutyn2014suicidal, phillips1974influence, stack2003media}, and there is no theoretical or practical reason that it would not also be found in acts of public violence. Confirming a contagion effect in public mass shootings would be of great value to those entrusted with protecting public safety.

Contagion theories have been developed since the late 19\textsuperscript{th} century across a variety of disciplines, with Bandura’s social learning theory proposing a cognitive framework through which attentive individuals learn aggression by observing the actions of others \cite{Bandura1973}. Bandura recognized that television and film broadcasts (and media attention in general) can serve as an effective transmission mechanism for aggressive behaviors (for a more complete discussion, see \cite{hornberger2025temporal}). And it is safe to assume that media attention would be the transmission mechanism for public mass shootings, as it is highly unlikely that any offender has personally observed a previous event \cite{meindl2017mass}.

Importantly, although all mass shootings receive significant media coverage, previous research has shown that public mass shootings receive far more coverage than non-public events (family mass shootings and mass shootings related to underlying criminal behavior), potentially making them particularly contagious \cite{fox2021newsworthiness, fox2021contagion, hornberger2025temporal}. Furthermore, the media attention directed to public mass shootings is overwhelmingly national, showing few regional differences, thus reducing the possibility of spatial effects of any contagion \cite{fox2021contagion}.

To date, contagion in public mass shooting has been understood in two ways. Copycatting or “specific contagion” occurs when an offender adopts a particular public mass shooter as a role model and attempts to carry out a similar act of violence \cite{langman2017role, langman2018different}. The authors in \cite{lankford2024similarities} recently documented evidence that a subset of mass shooters engages in copycat assaults (including statements made by the offenders themselves and striking demographic and behavioral parallels between offenders and role models) and demonstrated the degree of similarity between model and copycat. Notably, they found that the majority (nearly 80\%) of the offenders attacked more than a year after their role model with an average temporal gap of approximately eight years. 

In contrast, “general contagion” (also called “social contagion” \cite{hornberger2025temporal}) is more akin to the metaphor appropriated from epidemiology which  describes how behaviors spread across groups of people, and contemplates a closer temporal connection between events than does copycatting. While it has been suggested that a more apt term may be “generalized imitation” in which individuals learn to perform behaviors that are like those observed or described (although not always enacting an exact copy of the role model’s behavior) \cite{meindl2017mass}, the concept is the same -- the occurrence of an event makes it more likely that a similar event will occur in the near future. 

Support for general contagion of mass shootings is mixed. Some studies suggest a short-term effect, while others find no significant impact. \cite{towers2015contagion} examined both school shootings and mass killings using a self-excitation model, finding that incidents are temporally contagious for approximately two weeks. Replications of the Towers study using the same data failed to find evidence of contagion \cite{king2017random, lankford2018mass}, although Towers subsequently asserted that the discrepant findings result from the use of binned versus unbinned methods \cite{towers2018detecting}. More recently, \cite{fox2021contagion} used a multivariate point process to analyze public mass shootings, finding no evidence of an impact of media coverage on the short-term prevalence of events, and \cite{hornberger2025temporal} utilized a periodically-observed time-homogeneous Poisson process framework that failed to detect a prominent contagion effect for public mass killings (although there was some evidence of a subtle contagion effect).

Here, we take a slightly different approach based on the important but often overlooked point that public mass shootings are not uniform and that one key source of variation lies in the fate of the perpetrator. A substantial proportion of mass shooters exhibit suicidal intent or planning prior to the attack \cite{lankford2015mass}, indicating that expectations about survival are often embedded in the offender’s decision-making process. The perpetrator’s fate also shapes how long an incident remains visible. Prior research shows that public mass shootings that end in the assailant’s arrest rather than death receive substantially greater and longer-lasting media coverage \cite{fox2021newsworthiness}. This leads us to ask a central question: If a public mass shooter dies at the scene, does that make it more likely that the next offender will also die at the scene? And conversely, if a perpetrator survives, does that outcome influence future attackers to expect or aim for survival as well? In other words, does the fate of one perpetrator influence the fate of the next?

To be clear, we are not assessing the potential contagion of public mass shooting \textit{events} which have high barriers to completion (e.g., extensive planning, target selection, weapons acquisition) and where the time between inspiration and capability is often lengthy (potentially making them less susceptible to short-term contagion), but rather a single aspect of the phenomenon - the "fate" decision - which has lower barriers to completion (resolve by the offender) and where the time between inspiration and capability is potentially much shorter (as little as minutes to seconds).

In answering our questions, we use a bivariate Hawkes process i.e., a self-exciting point-process model introduced in \cite{hawkes1971spectra} to capture how the occurrence of one event increases the likelihood of the occurrence of another event in the near future. Hawkes models have been applied in a variety of domains, including seismology \cite{ogata1998space}, finance \cite{bacry2015hawkes}, epidemiology \cite{chiang2022hawkes, diouane2024accurate}, and criminology \cite{mohler2011self, leverso2025measuring}. Building on this framework, we fit a bivariate Hawkes process to public public mass shooting incidents in the United States from 1966 to 2024, where we distinguish between two categories. The first category, "die at the scene", includes perpetrators who end their own lives during the attack as well as those who are killed by law enforcement. The second category, which we refer to as "live", includes all remaining cases in which the perpetrator survives the immediate event and may later die from sustained injuries, by suicide, die in custody, be executed, flee, or remain alive in prison. In other words, "live" includes every outcome that is not death at the scene. 

The paper is organized as follows. In Section \ref{sec:method}, we provide a brief overview of point processes, introduce the bivariate Hawkes process, and describe the parameter estimation procedure. In Section \ref{sec:data} we provide detail about the data. Section \ref{sec:results} presents the results for the full 1966–2024 period as well as for the pre- and post-2000 eras. In Section \ref{sec:goodness of fit}, we assess the goodness of fit of the model. Section \ref{sec:discussion} discusses our findings, and Section \ref{sec:limitations} outlines the limitations of our approach and potential directions for future research.

\section{Method}
\label{sec:method}
\subsection{Point processes}
A point process is a random collection of occurrences of a certain event in time intervals or the number of points in regions of a space. Typical examples range from times of birth, police emergencies, public mass shootings, failures of machines, etc.

\begin{figure}[t]
\centering
\includegraphics[scale=0.5]{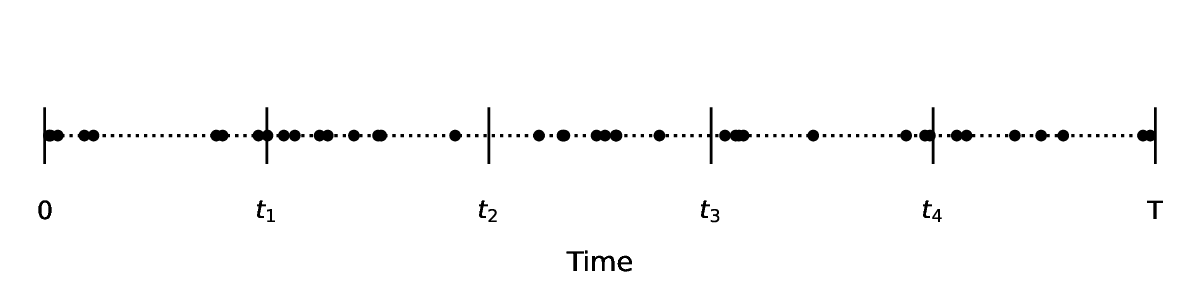}
\caption{An example of a simple temporal point process. The black bullets represent the occurrences of an event, e.g., 911 calls.}
\label{fig:temporal pp}
\end{figure}

Consider the case of a temporal point process denoted by  $\mathbf{N}$ (e.g., the times of occurrences of events between time 0 and $T$). One may describe $\mathbf{N}$ by a counting process $N(t)$, where for any time $t\in \{t_1, t_2, \dots, T\}$, $N(t)$ can be regarded as the number of points occurring at or before time $t$, $N(t)$ then reads:

\begin{equation}
N(t) = \sum\limits_{i = 1}^\infty \mathbf{1}_{\{t_i<t\}},
\end{equation}

\noindent where $\mathbf{1}$ represents the indicator function that takes a value 1 when $t_i < t$ and 0 otherwise. When modeling purely temporal data, the space in which the points fall is just a segment of the real line (see Figure \ref{fig:temporal pp}).

\begin{figure}
\centering
\includegraphics[scale=0.5]{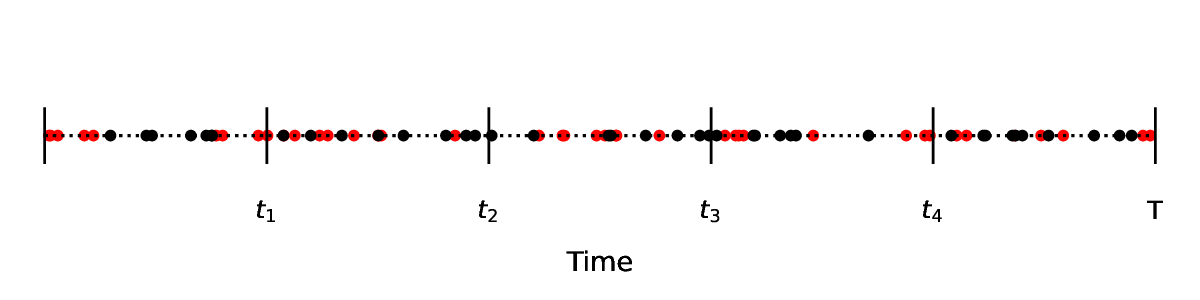}
\caption{An illustration of a marked temporal point process. The red bullets correspond to the occurrences of the events of type 1, while the black ones account for the events of type 2. An example could be gang violence online (black bullets) and offline (red bullets).}
\label{fig:temporal marked pp}
\end{figure}

\noindent It is worth mentioning that in addition to their associated times (and locations for the case of spatio-temporal point process), important information can be attributed to each point. For example, one may be interested in investigating the times of 911 calls along with their associated incidents, or the occurrences of social media posts of gang members and their associated crimes in real life. Such processes are known as marked point processes, where each point has a variable associated with it known as a mark or type. In Figure \ref{fig:temporal marked pp}, we show a marked temporal point process illustrating two event types.

\subsection{Univariate Hawkes processes}
\begin{figure}[t]
\centering
\includegraphics[scale=0.4]{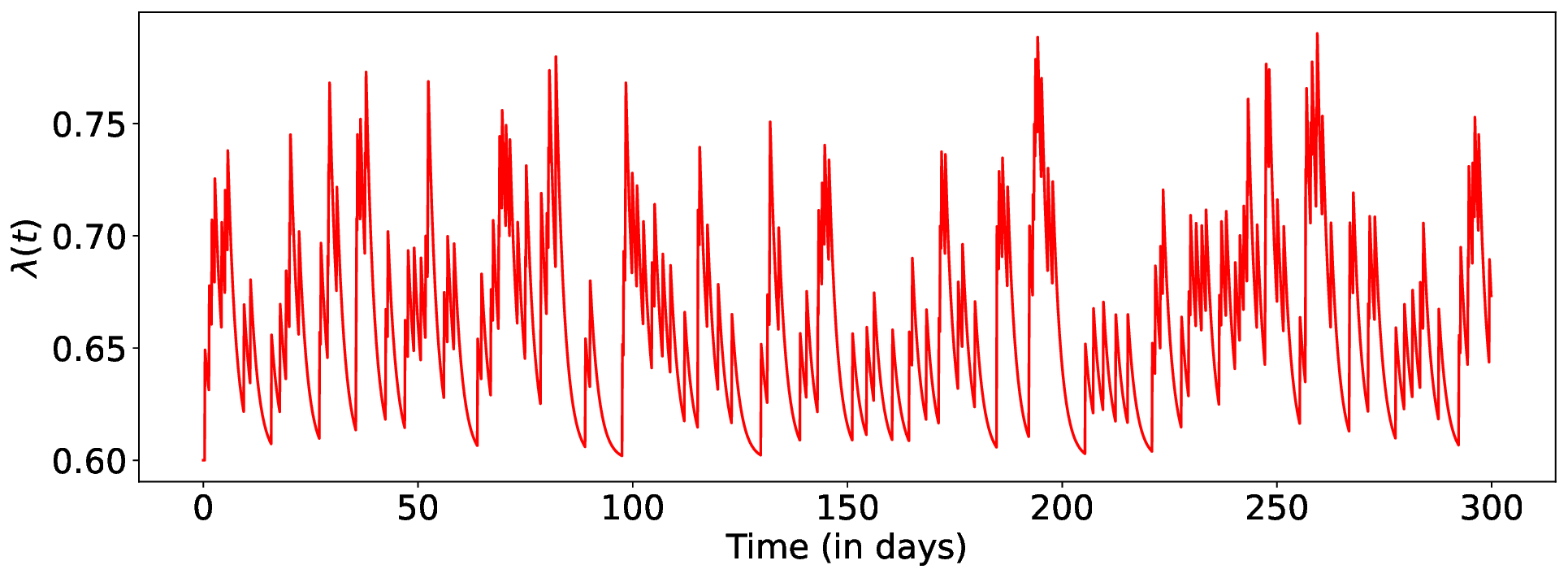}  
\caption{Example of a univariate Hawkes process conditional intensity using an exponential kernel. The parameters used in this simulation are, $\alpha = 0.1$, $\beta = 0.5$, $\mu = 0.6$ and $T = 300$.}
\label{fig:univeriate-hawkes}
\end{figure}

Consider a temporal simple point process of event times $\{t_1, t_2 \dots T\}$ such that $t_i < t_{i+1}$ and a continuous counting measure in $\mathbb{R}_+$ $N(A)$ defined as the number of events occurring at times $t\in A$. The process may be characterized by its conditional intensity \cite{reinhart2018review}

\begin{equation}
\lambda(t|H_t) = \lim_{\Delta t \rightarrow 0}\frac{\mathbb{E}\left[N(t, t + \Delta t)\right]}{\Delta t}
\end{equation} 

\noindent where $H_t$ represents the process's history of all events up to time $t$.  The Hawkes process \cite{hawkes1971spectra} is a self-exciting point process in which the occurrence of an event increases the probability of occurrence of another event in the near future. It can be defined in terms of a conditional intensity function in the equivalent form

\begin{equation}
\lambda_i(t) = \mu_i + \sum_{t_j < t} \varphi(t - t_j)
\end{equation}

where $\mu$ accounts for the baseline intensity that we consider time independent, and $\varphi$ is a non-negative function known as the triggering kernel, and models the increase in the intensity after the occurrence of event $i$ at time $t_i$. In the present study we use the exponential kernel $\varphi(t - t_i) = \alpha\beta e^{-\beta(t - t_i)}$. The model is called self-exciting because the current conditional intensity explicitly depends on the history $H_t$ of past events. Each event adds a contribution through the kernel function $\varphi(.)$, temporarily raising the current intensity and thereby increasing the likelihood of further events in the near future. Figure \ref{fig:univeriate-hawkes} illustrates a simulated univariate Hawkes process, with an exponential kernel. 

\subsection{Bivariate Hawkes processes}
While the univariate Hawkes process captures the self-exciting behavior of a single type of event, many real world systems involve interactions between multiple related event types. In the context of this study, we are interested in examining the interaction between two event types of public mass shooting outcomes:

\begin{itemize}
\item Type 1: incidents where the shooter dies at the scene ("die at the scene").
\item Type 2: incidents where the shooter lives("live").
\end{itemize} 

Type 1 refers to the case where the perpetrator does not survive beyond the immediate location and timeframe of the attack. This includes situations where the shooter is killed by law enforcement during the confrontation or dies by suicide on the site. Type 2 corresponds to situations where the perpetrator survives the immediate confrontation and later dies, either from injuries sustained during the event, suicide after leaving the scene, or through subsequent legal outcomes such as death in prison or execution.

\begin{figure}
\centering
\includegraphics[scale=0.4]{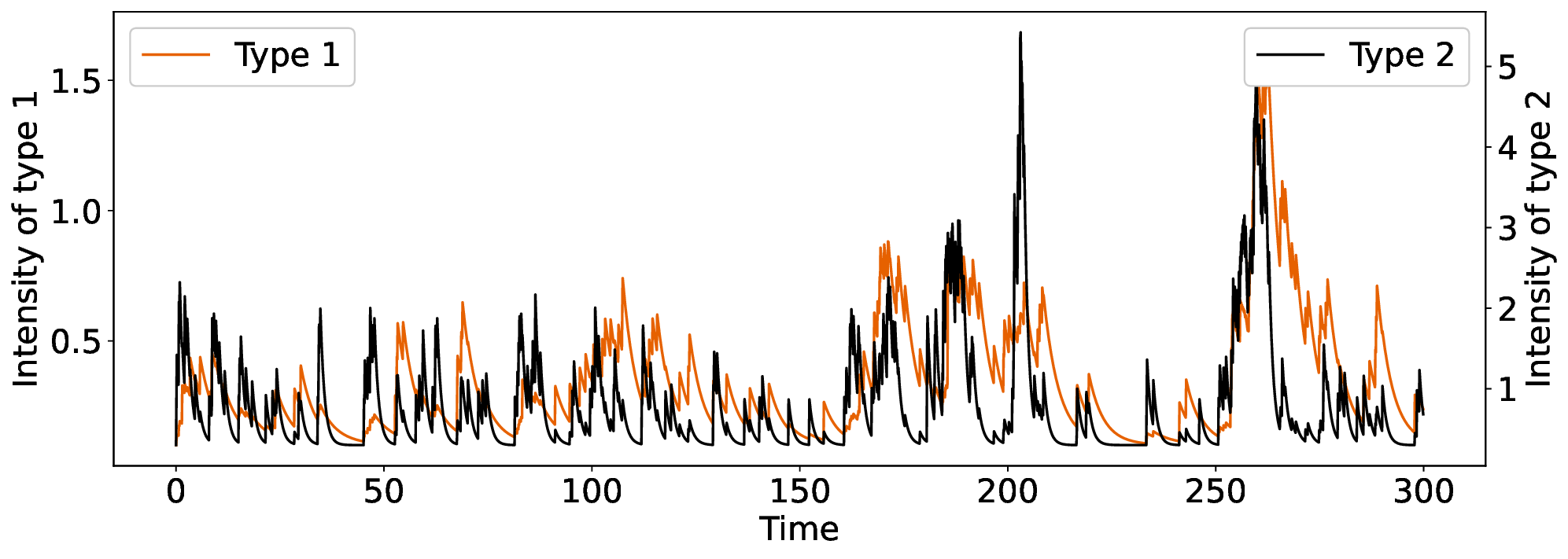}
\caption{Example of a bivariate Hawkes process conditional intensity. The parameters used are, $\alpha = \begin{pmatrix}
0.5 & 0.1 \\
0.2 & 0.6
\end{pmatrix}$, $\beta = \begin{pmatrix}
0.3 & 0.2 \\
0.8 & 1
\end{pmatrix}$, and $\mu = (0.1, 0.2)$ with $T = 300$ (in days).}
\label{fig:simulated-bivariate-hawkes}
\end{figure}

To capture the potential interaction between these two event types, we employ a bivariate Hawkes process, defined through the coupled conditional intensity

\begin{equation}
\lambda_i(t) = \mu_i + \sum\limits_{j = 1}^2 \sum_{k:t_k^j}^{n^j} \alpha_{ij}\beta_{ij}e^{-\beta_{ij}(t - t_k^j)}, \, \, \, \, i = 1,2
\label{eq:bivariate-hawkes}
\end{equation}

where $n^j$ represents the total number of events of type $j$, $\mu = (\mu_1, \mu_2)$ corresponds to the background rate of the incidents where the perpetrator dies at the scene and when they survive the attack (live), respectively, $\alpha$ is a $2\times 2$ matrix, known as the reproduction matrix whose elements $\alpha_{ij}$ describe the expected number of events of type $i$ that are directly triggered by one event of type $j$; and $\beta_{ij}$ corresponds to the decay of cross excitation from event $j$ to event $i$ and controls how quickly the excitation effect from an event of type $j$ on type $i$ decays over time. In Figure \ref{fig:simulated-bivariate-hawkes}, we show the coupled conditional intensity of a synthetic bivariate Hawkes process.

\subsection{Parameter estimation}
To estimate the model parameters, i.e., the reproduction matrix $\alpha$, the decay matrix $\beta$ and the baseline intensity $\mu$, we maximize the log-likelihood 

\begin{equation}
L = \sum_{i = 1}^2\sum\limits_{k = 1}^{n^j} \log(\lambda(t_k^i)) - \int_0^t \lambda(s)ds
\label{eq:log-likelihood}
\end{equation}

\noindent To evaluate the equation \eqref{eq:log-likelihood}, we need to analytically or approximately compute the integral $\int_0^s \lambda(s)ds$. We employ facilitated estimation \cite{schoenberg2013facilitated}, assuming that the observation time window $T$ is large compared to the longest timescale of the decay matrix $\beta$. Let $n^1$ and $n^2$ be the number of events of type 1 and 2, respectively, and $t_k^1$ and $t_k^2$ the event times of type 1 and 2, respectively. We have

\begin{equation}
\begin{split}
L &= \sum_{i = 1}^2\sum\limits_{k = 1}^{n^j} \log(\lambda(t_k^i)) - \sum_{j = 1}^2 \int_0^T \mu_j ds - 
\sum_{j = 1}^2\int_{t_k^1}^T\sum_{k:t_k^1}^{n^1}\alpha_{j1}\beta_{j1}e^{-\beta_{j1}(s-t_k^1)}ds\\
&\qquad - \sum_{j = 1}^2\int_{t_k^2}^T\sum_{k:t_k^2}^{n^2}\alpha_{j2}\beta_{j2}e^{-\beta_{j2}(s-t_k^2)}ds\\
&\simeq \sum_{i = 1}^2\sum\limits_{k = 1}^{n^j} \log(\lambda(t_k^i)) -\sum_{j=1}^2\mu_j T - \sum_{j=1}^2\left(\alpha_{j1} n^1 + \alpha_{j2}n^2\right)
\end{split}
\end{equation}

\noindent
Hence, the maximum log-likelihood can be approximated as

\begin{equation}
L \simeq \sum_{j = 1}^2 \left(\sum_{k = 1}^{n^j}\log\left(\lambda(t_k^j)\right) - \mu_j T - \alpha_{j1} n^1 - \alpha_{j2}n^2\right)
\label{eq:approximated-log-likelihood}
\end{equation}

\noindent We estimate the model parameters $\mu$, $\alpha$ and $\beta$ using the probabilistic language Stan \cite{carpenter2017stan}. To do so, we perform Hamiltonian Monte Carlo with 5,000 samples. The priors used are

\begin{equation}
\beta_{ij} \sim normal(0,1) \, \, , \, \, \alpha_{ij} \sim beta(1,1) \, \, , \, \, \mu_i \sim cauchy(0,5) \, \, \, \, i,j = 1,2
\end{equation} 

For the decay parameters $\beta$, we use a $normal(0,1)$ prior. This prior (constrained to be non-negative) is weakly informative. It encodes the expectation that contagion effects decay on the order of days to weeks \cite{towers2015contagion}, while still allowing slower or faster rates if supported by the data. For the excitation parameters $\alpha$, we specify $beta(1,1)$ priors, i.e., uniform distributions on the interval $(0,1)$, which enforce the theoretical constraint that excitation probabilities remain bounded between 0 and 1. Finally, for the baseline intensities $\mu$, we use $cauchy(0,5)$. This choice reflects the fact that we have no prior information about the parameter $\mu$, thus, we use an uninformative prior with high variance.

\section{Data}
\label{sec:data}
We obtained the data \footnote{\url{https://www.theviolenceproject.org/mass-shooter-database/}} from The Violent Prevention Project, a nonprofit research center dedicated to reducing violence through actionable research, which maintains a comprehensive open-source database on public mass shootings in the United States since 1966. The database defines public mass shooting as we do here - incidents where four or more people killed by firearms (excluding the offender), and disregarding cases primarily driven by gangs or drugs. The period covered for this study is from August 1966 to September 2024, with a total of 200  public mass shooting incidents across the U.S in which 1,438 people were killed. 

As mentioned previously, we classify these events into two categories: incidents in which the offender dies at the scene (128 cases) and incidents in which the shooter lives (72 cases). The model is then fitted to the data where the model parameters $\alpha$, $\beta$, and $\mu$ are estimated using the equation \eqref{eq:approximated-log-likelihood}.

\section{Results}
\label{sec:results}
\subsection{1966--2024}
\begin{figure}[t]
\centering
\includegraphics[scale=0.4]{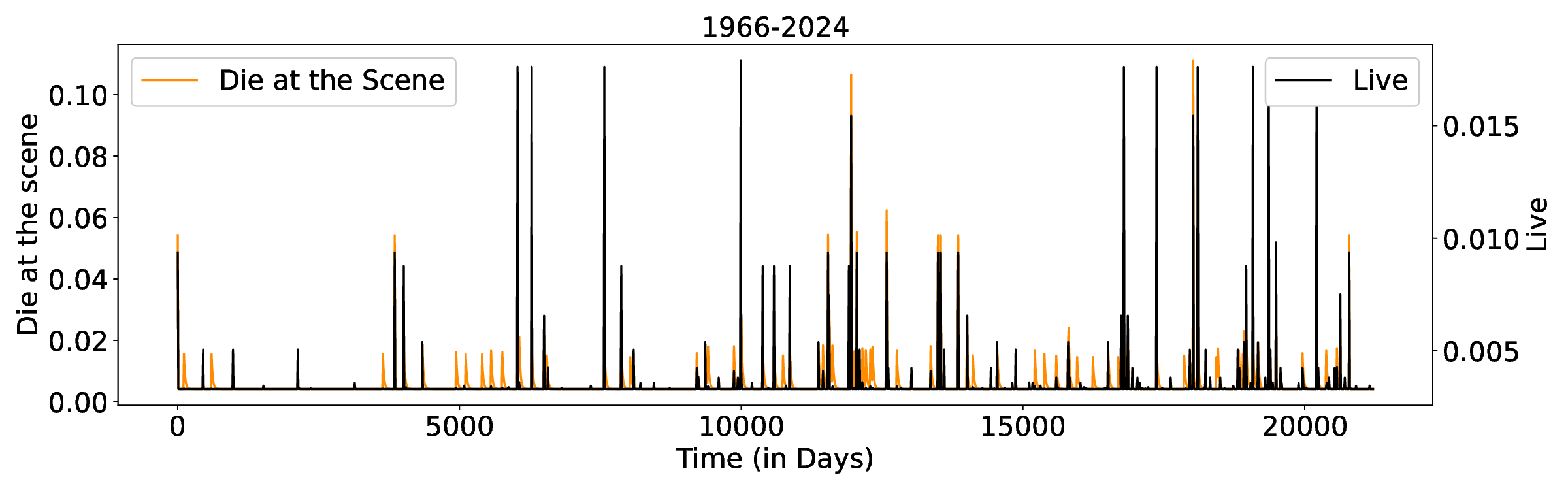}
\caption{Bivariate Hawkes process coupled conditional intensity for two event types: shooter dies at the scene (type 1) and shooter lives (type 2). The period covered is from August 1966 to September 2024, a total of 21,219 days.}
\label{fig:conditional-intensity}
\end{figure}

The results for the full period 1966--2024 regarding the parameter estimation of $\alpha$, $\beta$ and $\mu$ are reported in Tables \ref{table:alpha}, \ref{table:beta}, and \ref{table:mu}, respectively, while Figure \ref{fig:conditional-intensity} illustrates the coupled conditional intensity for the two event types. 

\begin{table}[t]
\centering
\begin{tabular}{|c|c|}
\hline
Reproduction matrix element & Mean (95\% confidence interval)\\
\hline
$\alpha_{11}$ & 0.139 (0.013, 0.528)\\
$\alpha_{12}$ & 0.343 (0.092, 0.806)\\
$\alpha_{21}$ & 0.018 (0.0007, 0.070)\\
$\alpha_{22}$ & 0.039 (0.004, 0.097)\\
\hline
\end{tabular}  
\caption{Estimated value of the reproduction matrix elements $\alpha$ as well as their corresponding 95\% confidence intervals for the period between 1966 and 2024. Index 1 corresponds to "die at the scene", while index 2 accounts for "live".} 
\label{table:alpha}
\end{table}

\begin{table}[t]
\centering
\begin{tabular}{|c|c|c|}
\hline
Decay rate & Mean (95\% confidence interval) & Time scale of contagion (in days)\\
\hline
$\beta_{11}$ & 1.226 (0.002, 2.972) & 0.816\\
$\beta_{12}$ &  0.050 (0.007, 0.112) & 20.161\\
$\beta_{21}$ & 0.621 (0.092, 1.748) & 1.610\\
$\beta_{22}$ & 0.963 (0.004, 2.250) & 1.038\\
\hline
\end{tabular}  
\caption{Estimated value of the decay matrix elements $\beta$ as well as their corresponding 95\% confidence intervals and the timescale of contagion in days for the period between 1966 and 2024. Index 1 corresponds to "die at the scene", while index 2 accounts for "live".} 
\label{table:beta}
\end{table} 

\begin{table}[t]
\centering
\begin{tabular}{|c|c|}
\hline
Baseline intensity & Mean (95\% confidence interval)\\
\hline
$\mu_1$ & 0.004 (0.002, 0.006)\\
$\mu_2$ & 0.003 (0.002, 0.004)\\
\hline
\end{tabular}  
\caption{Estimated values of the baseline rates for the two event types along with their corresponding 95\% confidence intervals. Index 1 corresponds to "die at the scene", while index 2 accounts for "live"}
\label{table:mu}
\end{table} 

From Table \ref{table:alpha}, we estimate $\alpha_{11} = 0.139$, which means that, on average, 0.139 of  public mass shooting events where the perpetrator dies at the scene are directly triggered by a single previous event of the same type. Furthermore, we estimate $\alpha_{22} = 0.039$, indicating that each event in which the perpetrator lives directly triggers, on average, 0.039 subsequent events of the same type. On the other hand, we estimate that $\alpha_{12} = 0.343$ of “die at the scene” events are directly caused by an initial “live” event (the strongest spillover effect). Finally, we observe that the lower bound of the 95\% confidence interval for $\alpha_{21} = 0.018$ is close to zero. Thus, the effect that each initial “die at the scene” event has on “live” events is not statistically significant.

The decay rate parameter matrix $\beta$ is given in Table \ref{table:beta}. Since $\beta$ is expressed in (days)$^{-1}$, we estimate the timescale of contagion from “die at the scene” to the same event type to be approximately 20 hours, while it is about one day for “live” to the same event type. The longest timescale of contagion is observed from “live” to “die at the scene,” lasting around 20 days, whereas the timescale from “die at the scene” to “live” is about one and a half day (though the effect of contagion was shown to be statistically not significant).

\subsection{Pre- and post-2000s}
In this subsection, we examine public mass shootings before and after the 2000s to investigate whether their underlying dynamics changed over time. The turn of the century marks an important transition point, characterized by social, technological, and cultural shifts that may have influenced both the frequency of incidents and the strength of contagion effects. Specifically, we compare the excitation effects and timescales of contagion between the two event types across the two periods. To this end, the dataset is divided into pre- and post-2000 (including the year 2000) subsets. The former includes 77 public mass shooting incidents, of which 41 involved perpetrators who died at the scene and 36 in which the shooter survived the attack, resulting in a total of 472 fatalities. The latter period comprises 123 incidents, where 87 offenders died at the scene and 36 survived the attack, accounting for 966 fatalities.

\begin{figure}
\centering
\includegraphics[scale=0.19]{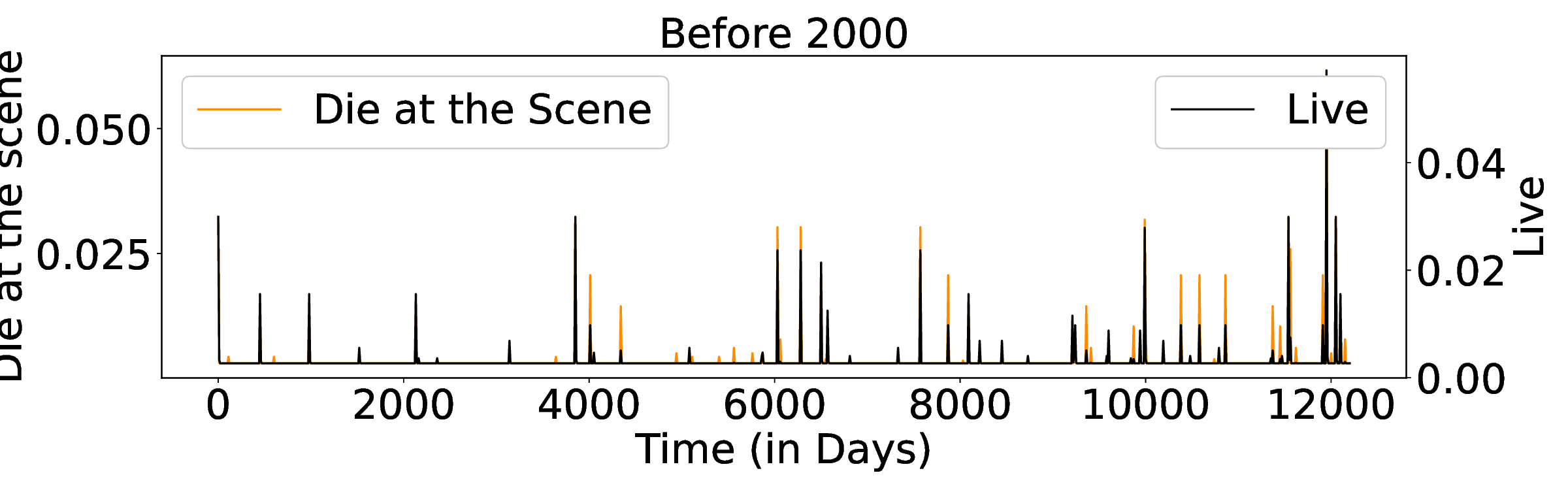} 
\includegraphics[scale=0.19]{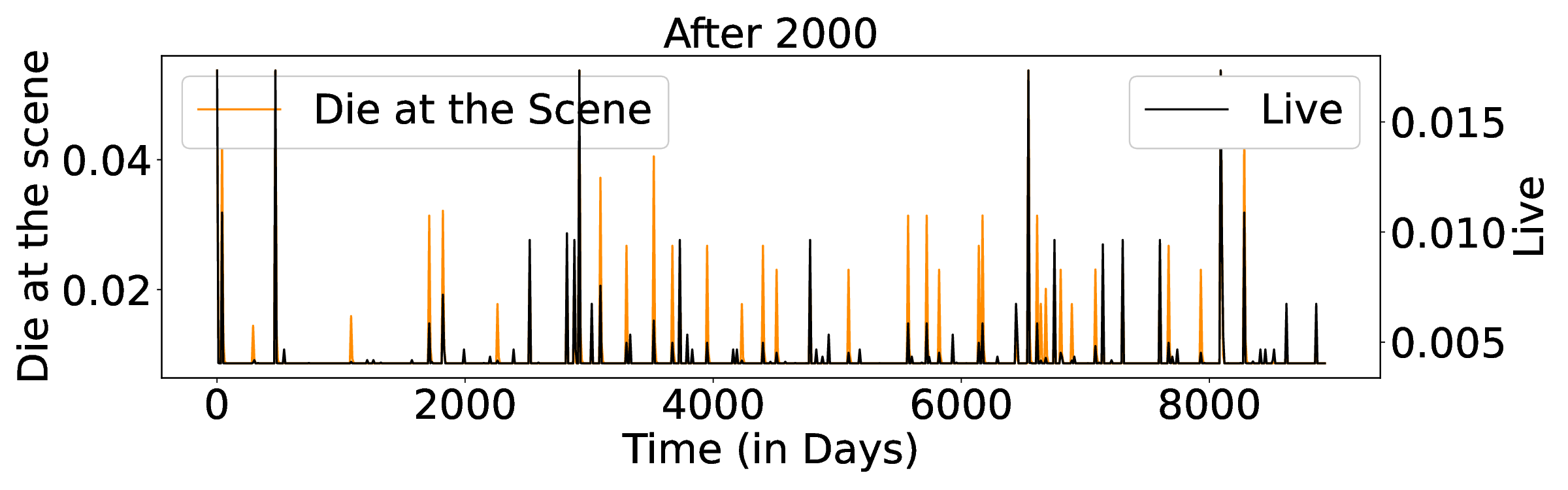}
\caption{Coupled intensity described in the equation \eqref{eq:bivariate-hawkes} before the year of 2000 (left) and after (right).}
\label{fig:conditional_intensity_2000}
\end{figure}

Next, we estimate the model parameters for the two periods and report the results in Tables \ref{table:alpha_2000}, \ref{table:beta_2000} and \ref{table:mu_2000}. Additionally, we plot the coupled conditional intensities for both periods in Figure \ref{fig:conditional_intensity_2000}.

\begin{table}[t]
\centering
\resizebox{\textwidth}{!}{
\begin{tabular}{|c|c|c|}
\hline
& Before 2000 & After 2000 \\
\hline
Reproduction matrix element& Mean (95\% confidence interval) & Mean (95\% confidence interval)\\
\hline
$\alpha_{11}$ & 0.089 (0.009, 0.229) & 0.044 (0.010, 0.091)\\
$\alpha_{12}$ & 0.097 (0.003, 0.348) & 0.248 (0.035, 0.550)\\
$\alpha_{21}$ & 0.106 (0.003, 0.367) & 0.016 (0.0007, 0.048)\\
$\alpha_{22}$ & 0.057 (0.007, 0.146) & 0.040 (0.002, 0.113)\\
\hline
\end{tabular}
}
\caption{Estimated values of the reproduction matrix elements as well as their corresponding 95\% confidence interval before and after the year 2000. Index 1 corresponds to "die at the scene", while index 2 accounts for "live".}  
\label{table:alpha_2000} 
\end{table}

\begin{table}[t]
\centering
\resizebox{\textwidth}{!}{
\begin{tabular}{|c|c|c|c|c|}
\hline
\multicolumn{1}{|c|}{} & \multicolumn{2}{c|}{Before 2000} & \multicolumn{2}{c|}{After 2000} \\
 \hline
$\beta$ & Mean (95\% CI) & Timescale of contagion (in days) & Mean (95\% CI) & Timescale of contagion (in days)\\
\hline
$\beta_{11}$ & 0.594 (0.048, 1.975) & 1.682 & 1.489 (0.347, 2.760) & 0.671 \\
$\beta_{12}$ & 0.434 (0.006, 1.649) & 2.302 & 0.228 (0.023, 0.960) & 4.382 \\
$\beta_{21}$ & 0.374 (0.004, 1.590) & 2.671 & 0.730 (0.029, 1.822) & 1.370 \\
$\beta_{22}$ & 1.087 (0.092, 2.379) & 0.920 & 0.664 (0.025, 1.774) & 1.505\\
\hline
\end{tabular}
}
\caption{Estimated values of the decay rate matrix elements and their corresponding 95\% confidence intervals before and after the year 2000, along with the associated timescales of contagion (in days). Index 1 corresponds to "die at the scene", while index 2 accounts for "live"}
\label{table:beta_2000} 
\end{table}

\begin{table}[t]
\centering
\begin{tabular}{|c|c|c|}
\hline
 & Before 2000 & After 2000 \\
 \hline
Baseline rate & Mean (95\% confidence interval) & Mean (95\% confidence interval)\\
\hline
$\mu_1$ & 0.009 (0.007, 0.011) & 0.003 (0.002, 0.004)\\
$\mu_2$ & 0.004 (0.003, 0.005) & 0.003 (0.002, 0.004)\\
\hline
\end{tabular}
\caption{Estimated values of the baseline rates for the two event types, along with their corresponding 95\% confidence intervals, before and after the year 2000. Index 1 corresponds to "die at the scene", while index 2 accounts for "live".}  
\label{table:mu_2000} 
\end{table}

From Table \ref{table:alpha_2000}, we notice that before 2000, the cross excitation effects i.e., "live"$\leftrightarrow$"die at the scene" are of the same magnitude, with $\alpha_{12} \simeq \alpha_{21} \simeq 0.1$, whereas after the year 2000, the effect "live"$\rightarrow$"die at the scene" significantly increased to $\alpha_{12} = 0.248$, while the "die at the scene"$\rightarrow$"live" decreased to $\alpha_{21} = 0.016$ with the lower end of the 95\% confidence interval being nearly 0, which indicates that the effect is not statistically significant. On the other hand, we note a decrease in the self-excitation effect, from $\alpha_{11} = 0.089$ before the year 2000 to $\alpha_{11} = 0.044$ after, and from $\alpha_{22} = 0.057$ before to $\alpha_{22} = 0.04$ after. 

The estimated timescales of contagion show clear differences before and after 2000. Before 2000, the timescale of contagion from "die at the scene" to "live" and from "live" to "die at the scene" is approximately the same, i.e., two days and a half. After 2000, the contagion timescale increased to 4.38 days from “live” to “die at the scene,” and decreased to 1.37 days in the opposite direction (although the spillover in this direction is not statistically significant). For self-excitation, the “die at the scene”$\rightarrow$“die at the scene” timescale of contagion shortened from 1.68 to 0.67 days (about 16 hours), while the one for “live”$\rightarrow$“live” increased from 0.92 (about 22 hours) to one day and a half.

\section{Goodness of fit}
\label{sec:goodness of fit}
In order to assess the goodness of fit of the model, we use residual analysis \cite{ogata1988statistical}. Consider the rescaled event times 

\begin{equation}
\tau_i = \int_0^{t_i} \lambda(s)ds.
\end{equation} 

\begin{figure}[t]
\centering
\includegraphics[scale=0.3]{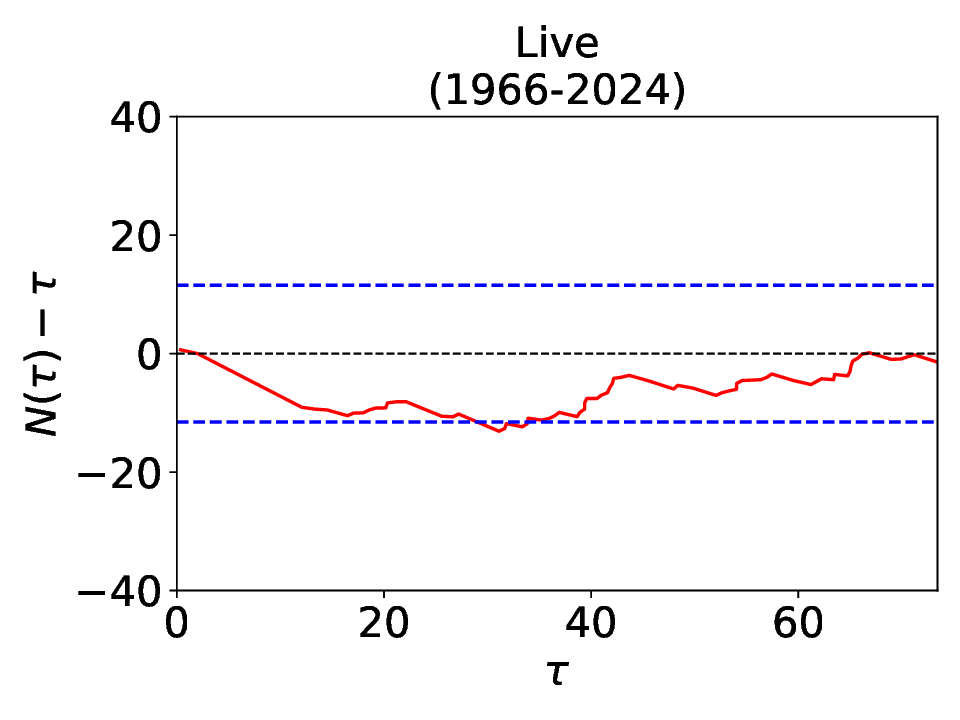} \includegraphics[scale=0.3]{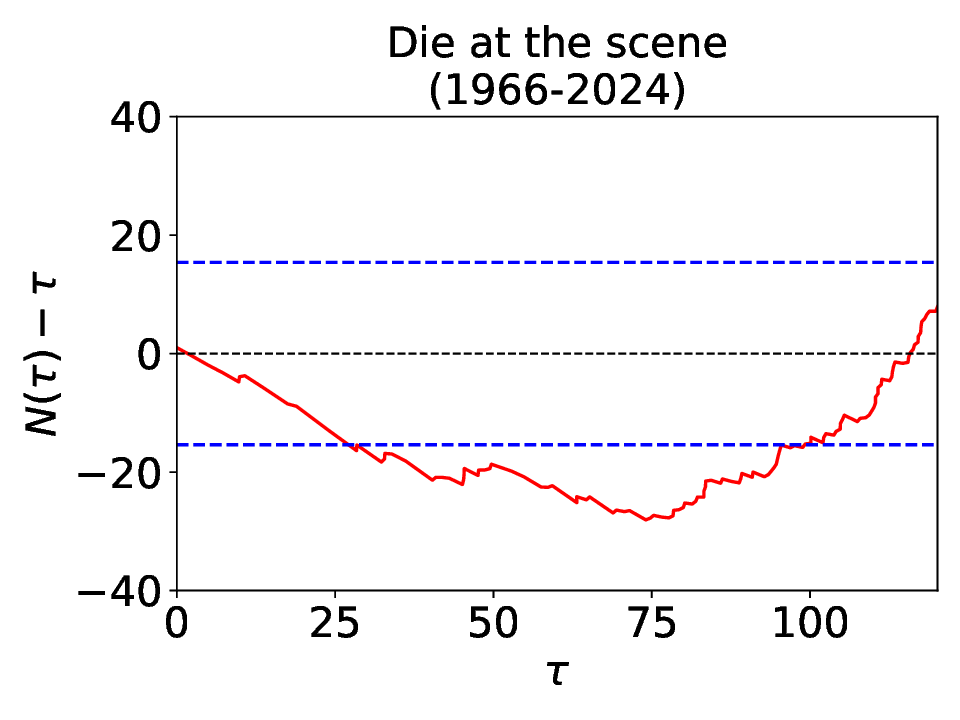} \includegraphics[scale=0.3]{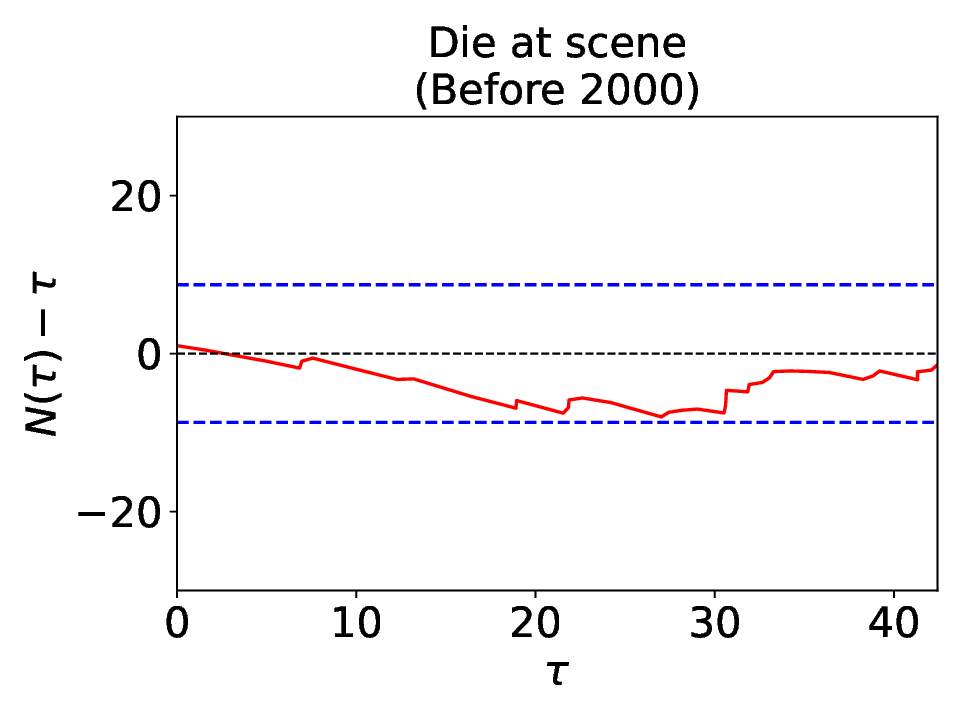} \includegraphics[scale=0.3]{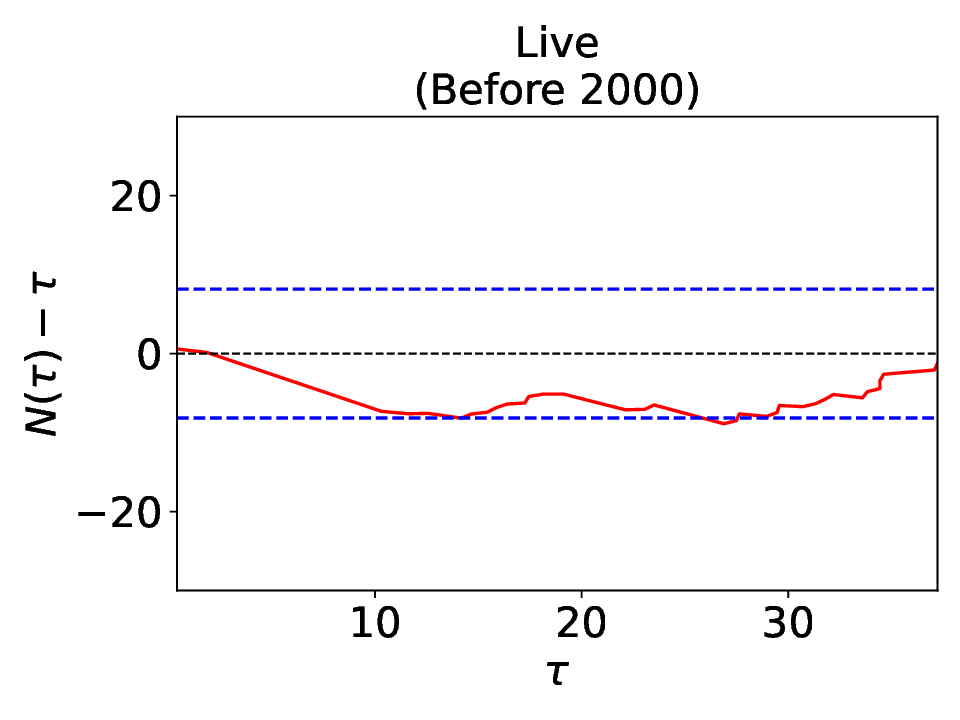} \includegraphics[scale=0.3]{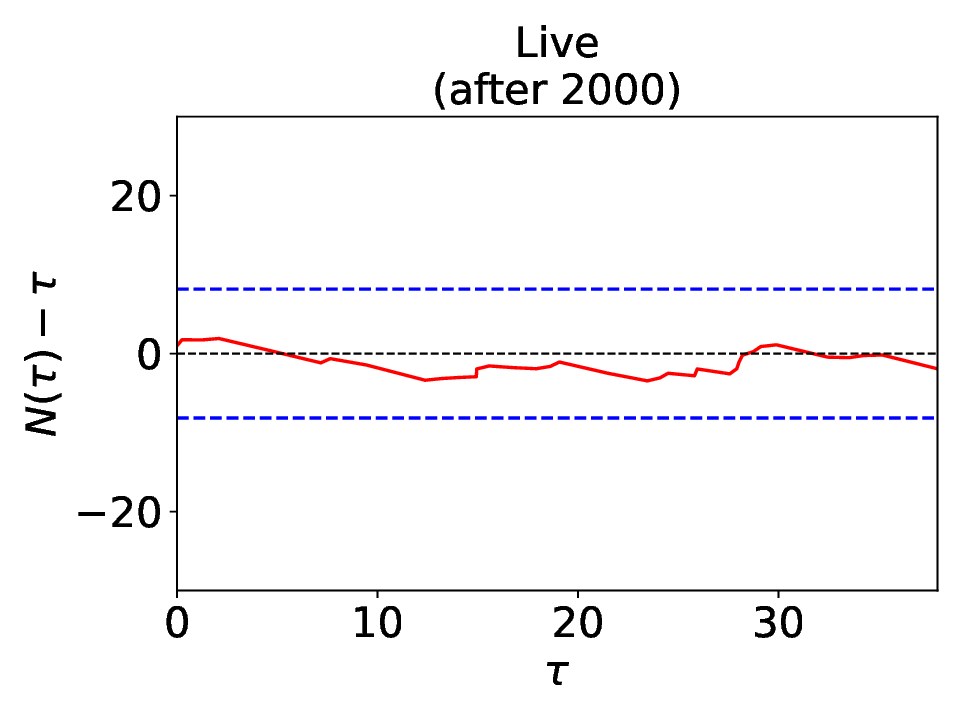} \includegraphics[scale=0.3]{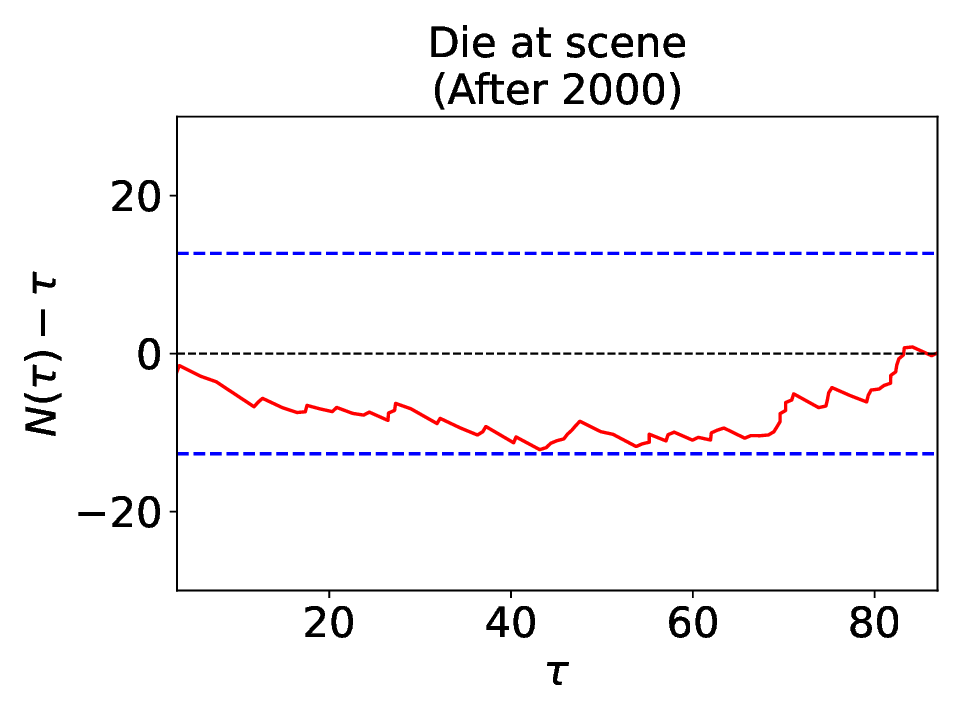} 
\caption{Normalized cumulative count along with the 95\% KS error bounds of the rescaled event times for the different periods, i.e., 1966--2024, before 2000 and after 2000. The dotted blue lines represent the KS bounds, while the red curve account for the rescaled data.}
\label{fig:goodness_of_fit}
\end{figure}

\noindent If the estimated conditional intensity is a good approximation to the true conditional intensity, then the transformed data $\{\tau_i\}$ are expected to be distributed according to a unit Poisson process. In Figure \ref{fig:goodness_of_fit} we plot the normalized cumulative number of events $N(\tau) - \tau$ against the rescaled event times along with the 95\% error bounds of the Kolmogorov-Smirnov (KS) statistic \cite{mohler2013modeling} for the datasets spanning the periods 1966--2024, before the year 2000, and after the year 2000. 

From Figure \ref{fig:goodness_of_fit}, we see that the estimated intensity provides a good fit for the “Live” data across the three studied periods, as illustrated by the rescaled event times remaining within the 95\% KS confidence bounds. The model also fits the “Die at the scene” data well for the periods before and after 2000; however, when the entire 1966--2024 dataset is modeled together, we observe a deviation from the 95\% KS bounds. This deviation suggests that the underlying process is not stationary over the full time span, meaning that there is a shift in the dynamics of public mass shootings where the offender dies at the scene from pre-2000 to post-2000. When the data are divided into pre-2000 and post-2000 intervals, these structural differences are isolated, allowing the model to better capture the temporal behavior within each period.

\section{Discussion}
\label{sec:discussion}

In this study, we provided new insight into the dynamics of public mass shootings in the United States by distinguishing between two types of incidents: those in which the perpetrator dies at the scene, either by suicide or as a result of police intervention, and those in which the perpetrator lives. Our objective was to examine how these two categories of events influence each other. In other words, we wanted to understand the spillover effects that each type of public mass shooting exerts on the other. To achieve this, we fitted a bivariate Hawkes process to public mass shooting data from 1966 to 2024. To further explore how these dynamics may have evolved over time, we divided the dataset into two sub-periods, before and after the year 2000 and estimated the model parameters separately for each period.

When analyzing the data for the period 1966--2024, we estimated that on average, 0.343 of “die at the scene” incidents are directly triggered by a previous “live” event, representing the strongest spillover effect. In contrast, only 0.018 of “live” incidents are directly caused by an initial “die at the scene” event, an effect that is weak and not statistically significant, with the lower bound of the 95\% confidence interval approaching zero. Additionally, we estimated that, on average, 0.139 of “die at the scene” incidents are directly triggered by the occurrence of a preceding event of the same type, whereas  0.039 of “live” incidents are caused by a prior “live” event. These findings indicate that “die at the scene” public mass shootings can only cause "die at the scene" public mass shootings, while “live” incidents act as the main source of contagion causing both “die at the scene” (highest spillover effect) and "live" events.

Why does the strongest spillover run from "live" to "die at the scene" and not the other way around? The answer may be in how these types of events are processed socially and psychologically. "Live" cases likely remain in the public view for longer due to investigations, court proceedings, and retrospective reporting, which amplifies their resonance and provides detailed narratives on motives and methods \cite{surette2016measuring, meindl2017mass}. This is consistent with the fact that "live"$\rightarrow$"die at the scene" has the longest timescale of contagion among all other types of influence, i.e., 20 days. The next question is why the most impacted event type from "live" is "die at the scene" and not "live"? A possible explanation is that the potential imitators, through expecting their act to end their life (by suicide or by resisting arrest and confronting the police so that they end being killed)  want to achieve notoriety without enduring the loss of control, humiliation, and prolonged scrutiny associated with survival. The finding that “die at the scene” incidents primarily trigger subsequent “die at the scene” events reinforces this interpretation. Their media trajectory is intense but short-lived. Indeed, as shown in Table \ref{table:beta}, the shortest timescale of contagion among all interactions corresponds to the “die at the scene”$\rightarrow$“die at the scene” pathway (approximately 20 hours). This brief period of heightened attention constrains public exposure and thus limits the potential of such events to generate “live” incidents. The contagion they produce appears to operate through direct behavioral imitation, i.e., the expectation of death. In contrast, events in which the perpetrator survives seem to propagate through symbolic imitation, i.e., the replication of the act itself.

To understand the public mass shooting dynamics even more deeply, we examined how these contagion mechanisms changed over time, notably before and after the year 2000. The year 2000 represents a natural dividing line for this purpose. Around that time, several key transformations occurred in the United States (and the world), such as the expansion of 24-hour cycle news coverage, the emergence of internet-based reporting, and the early stages of social media. These developments reshaped how violent events were publicized and consumed \cite{thompson2019media, girgis2025changes}, potentially modifying both the frequency of incidents and the strength of contagion between them. Our results indicate that prior to 2000, the cross-excitation effects between the two event types were approximately equal, with $\alpha_{12} \simeq \alpha_{21} \simeq 0.1$, and the corresponding contagion timescales were also similar, about two and a half days for both interactions. After 2000, however, the triggering effect of each “live” event on “die at the scene” incidents increased substantially to $\alpha_{12} = 0.2$, while the reverse effect became statistically not significant, consistent with the dynamics observed over the full 1966 to 2024 period. This suggests that before 2000, public mass shootings where the perpetrator died or survived were perceived and processed similarly, resulting in cross-excitation effects of approximately similar magnitudes. After 2000, the dynamics shifted toward the pattern seen in the full period, with the strongest spillover running from “live” to “die at the scene,” accompanied by the longest contagion timescale of about four and a half days. The reverse effect lost significance, and “die at the scene” incidents were found to trigger only events of the same type, characterized by the shortest contagion timescale of about 15 hours. This shift likely reflects the profound changes in media dynamics and public engagement that accompanied the digital era. After 2000, the visibility of “live” perpetrators through televised trials, online manifestos, and continuous digital commentary extended far beyond the initial event. Such prolonged exposure reinforced the symbolic presence of these individuals and provided detailed narratives that potential imitators could identify with or seek to emulate. The persistence of  the “die at the scene” self-excitation pattern, already observed in the full-period analysis, further supports this interpretation. Its influence became even more temporally constrained after 2000, suggesting that while these incidents continue to trigger public mass shootings of similar type, their broader contagion potential has diminished over time.


\section{Limitations and future directions}
\label{sec:limitations}
One limitation of this study is that public mass shootings are extremely rare events, with only 200 incidents recorded over 58 years. The scarcity of data increases the risk of overfitting and limits the robustness of parameter estimates, making regularization techniques necessary to ensure stable inference \cite{salehi2019learning}. Future research could address this limitation by expanding the analysis to include public mass shootings worldwide, thereby improving statistical power and enabling broader generalization. Another limitation is that the model focuses exclusively on temporal dependencies and does not incorporate covariates such as geography, gender, media exposure, socioeconomic context, or the psychological state of perpetrators prior to the act. Integrating these covariates into a Hawkes process could provide a more comprehensive understanding of the mechanisms driving contagion. Finally, this study assumes an exponential kernel to model the excitation effect. A nonparametric Hawkes process that learns the kernel shape directly from the data could offer a more flexible representation of event dynamics \cite{boyd2021assessing}. However, such an approach would require substantially more data \cite{malem2022variational} than currently available for public mass shootingss. Further research exploring this direction would be valuable as larger and more detailed datasets become available.

\bibliographystyle{vancouver}
\bibliography{bibitem}

@article{reinhart2018review,
  title={A review of self-exciting spatio-temporal point processes and their applications},
  author={Reinhart, Alex},
  journal={Statistical Science},
  volume={33},
  number={3},
  pages={299--318},
  year={2018},
  publisher={JSTOR}
}

@article{hawkes1971spectra,
  title={Spectra of some self-exciting and mutually exciting point processes},
  author={Hawkes, Alan G},
  journal={Biometrika},
  volume={58},
  number={1},
  pages={83--90},
  year={1971},
  publisher={Oxford University Press}
}

@article{ogata1998space,
  title={Space-time point-process models for earthquake occurrences},
  author={Ogata, Yosihiko},
  journal={Annals of the Institute of Statistical Mathematics},
  volume={50},
  number={2},
  pages={379--402},
  year={1998},
  publisher={Springer}
}

@article{mohler2011self,
  title={Self-exciting point process modeling of crime},
  author={Mohler, George O and Short, Martin B and Brantingham, P Jeffrey and Schoenberg, Frederic Paik and Tita, George E},
  journal={Journal of the american statistical association},
  volume={106},
  number={493},
  pages={100--108},
  year={2011},
  publisher={Taylor \& Francis}
}

@article{chiang2022hawkes,
  title={Hawkes process modeling of COVID-19 with mobility leading indicators and spatial covariates},
  author={Chiang, Wen-Hao and Liu, Xueying and Mohler, George},
  journal={International journal of forecasting},
  volume={38},
  number={2},
  pages={505--520},
  year={2022},
  publisher={Elsevier}
}

@article{carpenter2017stan,
  title={Stan: A probabilistic programming language},
  author={Carpenter, Bob and Gelman, Andrew and Hoffman, Matthew D and Lee, Daniel and Goodrich, Ben and Betancourt, Michael and Brubaker, Marcus and Guo, Jiqiang and Li, Peter and Riddell, Allen},
  journal={Journal of statistical software},
  volume={76},
  pages={1--32},
  year={2017}
}

@article{schoenberg2013facilitated,
  title={Facilitated estimation of ETAS},
  author={Schoenberg, Frederic Paik},
  journal={Bulletin of the seismological Society of America},
  volume={103},
  number={1},
  pages={601--605},
  year={2013},
  publisher={Seismological Society of America}
}

@article{leverso2025measuring,
  title={Measuring Online--Offline Spillover of Gang Violence Using Bivariate Hawkes Processes},
  author={Leverso, John and Diouane, Youness and Mohler, George},
  journal={Journal of quantitative criminology},
  volume={41},
  number={1},
  pages={103--131},
  year={2025},
  publisher={Springer}
}

@article{towers2015contagion,
  title={Contagion in mass killings and school shootings},
  author={Towers, Sherry and Gomez-Lievano, Andres and Khan, Maryam and Mubayi, Anuj and Castillo-Chavez, Carlos},
  journal={PLoS one},
  volume={10},
  number={7},
  pages={e0117259},
  year={2015},
  publisher={Public Library of Science}
}

@article{ogata1988statistical,
  title={Statistical models for earthquake occurrences and residual analysis for point processes},
  author={Ogata, Yosihiko},
  journal={Journal of the American Statistical association},
  volume={83},
  number={401},
  pages={9--27},
  year={1988},
  publisher={Taylor \& Francis}
}

@article{mohler2013modeling,
  title={Modeling and estimation of multi-source clustering in crime and security data},
  author={Mohler, George},
  journal={The Annals of Applied Statistics},
  pages={1525--1539},
  year={2013},
  publisher={JSTOR}
}

@article{surette2016measuring,
  title={Measuring copycat crime},
  author={Surette, Ray},
  journal={Crime, media, culture},
  volume={12},
  number={1},
  pages={37--64},
  year={2016},
  publisher={Sage Publications Sage UK: London, England}
}

@article{meindl2017mass,
  title={Mass shootings: The role of the media in promoting generalized imitation},
  author={Meindl, James N and Ivy, Jonathan W},
  journal={American journal of public health},
  volume={107},
  number={3},
  pages={368--370},
  year={2017},
  publisher={American Public Health Association}
}

@article{thompson2019media,
  title={Media exposure to mass violence events can fuel a cycle of distress},
  author={Thompson, Rebecca R and Jones, Nickolas M and Holman, E Alison and Silver, Roxane Cohen},
  journal={Science advances},
  volume={5},
  number={4},
  pages={eaav3502},
  year={2019},
  publisher={American Association for the Advancement of Science}
}

@article{girgis2025changes,
  title={Changes in Rates of Suicide by Mass Shooters, 1980--2019},
  author={Girgis, Ragy R and Hesson, Hannah and Brucato, Gary and Lieberman, Jeffrey A and Appelbaum, Paul S and Mann, J John},
  journal={Archives of suicide research},
  volume={29},
  number={1},
  pages={317--326},
  year={2025},
  publisher={Taylor \& Francis}
}

@article{salehi2019learning,
  title={Learning hawkes processes from a handful of events},
  author={Salehi, Farnood and Trouleau, William and Grossglauser, Matthias and Thiran, Patrick},
  journal={Advances in neural information processing systems},
  volume={32},
  year={2019}
}

@article{boyd2021assessing,
  title={Assessing the contagiousness of mass shootings with nonparametric Hawkes processes},
  author={Boyd, Peter and Molyneux, James},
  journal={PLoS one},
  volume={16},
  number={3},
  pages={e0248437},
  year={2021},
  publisher={Public Library of Science San Francisco, CA USA}
}

@article{malem2022variational,
  title={Variational Bayesian inference for nonlinear Hawkes process with Gaussian process self-effects},
  author={Malem-Shinitski, Noa and Ojeda, C{\'e}sar and Opper, Manfred},
  journal={Entropy},
  volume={24},
  number={3},
  pages={356},
  year={2022},
  publisher={MDPI}
}

@article{lankford2015mass,
  title={Mass murderers in the United States: Predictors of offender deaths},
  author={Lankford, Adam},
  journal={The Journal of Forensic Psychiatry \& Psychology},
  volume={26},
  number={5},
  pages={586--600},
  year={2015},
  publisher={Taylor \& Francis}
}

@article{bacry2015hawkes,
  title={Hawkes processes in finance},
  author={Bacry, Emmanuel and Mastromatteo, Iacopo and Muzy, Jean-Fran{\c{c}}ois},
  journal={Market Microstructure and Liquidity},
  volume={1},
  number={01},
  pages={1550005},
  year={2015},
  publisher={World Scientific}
}

@inproceedings{diouane2024accurate,
  title={Accurate estimation of cross-excitation in multivariate Hawkes process models of infectious diseases},
  author={Diouane, Youness and Schoenberg, Frederic and Mohler, George},
  booktitle={2024 IEEE 11th International Conference on Data Science and Advanced Analytics (DSAA)},
  pages={1--8},
  year={2024},
  organization={IEEE}
}

@misc{krouse2015mass,
  title={Mass murder with firearms: Incidents and victims, 1999-2013},
  author={Krouse, William J and Richardson, Daniel J},
  year={2015},
  publisher={Congressional Research Service Washington, DC}
}

@article{chermak1998predicting,
  title={Predicting crime story salience: The effects of crime, victim, and defendant characteristics},
  author={Chermak, Steven},
  journal={Journal of Criminal Justice},
  volume={26},
  number={1},
  pages={61--70},
  year={1998},
  publisher={Elsevier}
}

@article{duwe2000body,
  title={Body-count journalism: The presentation of mass murder in the news media},
  author={Duwe, Grant},
  journal={Homicide Studies},
  volume={4},
  number={4},
  pages={364--399},
  year={2000},
  publisher={Sage Publications, Inc.}
}

@article{duwe2004patterns,
  title={The patterns and prevalence of mass murder in twentieth-century America},
  author={Duwe, Grant},
  journal={Justice Quarterly},
  volume={21},
  number={4},
  pages={729--761},
  year={2004},
  publisher={Taylor \& Francis}
}

@article{fox2021newsworthiness,
  title={The newsworthiness of mass public shootings: What factors impact the extent of coverage?},
  author={Fox, James Alan and Gerdes, Madison and Duwe, Grant and Rocque, Michael},
  journal={Homicide studies},
  volume={25},
  number={3},
  pages={239--255},
  year={2021},
  publisher={SAGE Publications Sage CA: Los Angeles, CA}
}

@article{fridel2022integrating,
  title={Integrating the literature on lethal violence: A comparison of mass murder, homicide, and homicide-suicide},
  author={Fridel, Emma E},
  journal={Homicide studies},
  volume={26},
  number={2},
  pages={123--147},
  year={2022},
  publisher={SAGE Publications Sage CA: Los Angeles, CA}
}

@article{schildkraut2016mass,
  title={Mass shootings: Media, myths, and realities: Media, myths, and realities (ABC},
  author={Schildkraut, J and Elsass, HJ},
  journal={CLIO},
  year={2016}
}

@article{silver2022sequence,
  title={A sequence analysis of the behaviors and experiences of the deadliest public mass shooters},
  author={Silver, James and Silva, Jason R},
  journal={Journal of interpersonal violence},
  volume={37},
  number={23-24},
  pages={NP23468--NP23494},
  year={2022},
  publisher={SAGE Publications Sage CA: Los Angeles, CA}
}

@article{fox2021contagion,
  title={The contagion of mass shootings: The interdependence of large-scale massacres and mass media coverage},
  author={Fox, James Alan and Sanders, Nathan E and Fridel, Emma E and Duwe, Grant and Rocque, Michael},
  journal={Statistics and Public Policy},
  volume={8},
  number={1},
  pages={53--66},
  year={2021},
  publisher={Taylor \& Francis}
}

@article{holden1986contagiousness,
  title={The contagiousness of aircraft hijacking},
  author={Holden, Robert T},
  journal={American Journal of Sociology},
  volume={91},
  number={4},
  pages={874--904},
  year={1986},
  publisher={University of Chicago Press}
}

@article{christakis2008collective,
  title={The collective dynamics of smoking in a large social network},
  author={Christakis, Nicholas A and Fowler, James H},
  journal={New England journal of medicine},
  volume={358},
  number={21},
  pages={2249--2258},
  year={2008},
  publisher={Mass Medical Soc}
}

@article{abrutyn2014suicidal,
  title={Are suicidal behaviors contagious in adolescence? Using longitudinal data to examine suicide suggestion},
  author={Abrutyn, Seth and Mueller, Anna S},
  journal={American sociological review},
  volume={79},
  number={2},
  pages={211--227},
  year={2014},
  publisher={Sage Publications Sage CA: Los Angeles, CA}
}

@article{phillips1974influence,
  title={The influence of suggestion on suicide: Substantive and theoretical implications of the Werther effect},
  author={Phillips, David P},
  journal={American sociological review},
  pages={340--354},
  year={1974},
  publisher={JSTOR}
}

@article{stack2003media,
  title={Media coverage as a risk factor in suicide},
  author={Stack, Steven},
  journal={Journal of Epidemiology \& Community Health},
  volume={57},
  number={4},
  pages={238--240},
  year={2003},
  publisher={BMJ Publishing Group Ltd}
}

@book{Bandura1973,
  author = {Bandura, Albert},
  title = {Aggression: A Social Learning Analysis},
  publisher = {Prentice-Hall},
  year = {1973},
  address = {Englewood Cliffs, NJ}
}

@article{hornberger2025temporal,
  title={Temporal analysis of the clustering and hypothesized social contagion of mass killing events in the United States},
  author={Hornberger, Zachary T and King, Douglas M and Jacobson, Sheldon H},
  journal={Socio-Economic Planning Sciences},
  pages={102349},
  year={2025},
  publisher={Elsevier}
}

@article{langman2017role,
  title={Role models, contagions, and copycats: An exploration of the influence of prior killers on subsequent attacks},
  author={Langman, Peter},
  journal={Available on www. schoolshooters. info},
  year={2017}
}

@article{langman2018different,
  title={Different types of role model influence and fame seeking among mass killers and copycat offenders},
  author={Langman, Peter},
  journal={American Behavioral Scientist},
  volume={62},
  number={2},
  pages={210--228},
  year={2018},
  publisher={Sage Publications Sage CA: Los Angeles, CA}
}

@article{lankford2024similarities,
  title={Similarities between copycat mass shooters and their role models: An empirical analysis with implications for threat assessment and violence prevention},
  author={Lankford, Adam and Silva, Jason R},
  journal={Journal of Criminal Justice},
  volume={95},
  pages={102316},
  year={2024},
  publisher={Elsevier}
}

@article{king2017random,
  title={Random acts of violence? Examining probabilistic independence of the temporal distribution of mass killing events in the United States},
  author={King, Douglas M and Jacobson, Sheldon H},
  journal={Urbana},
  volume={51},
  pages={61801},
  year={2017}
}

@article{lankford2018mass,
  title={Mass killings in the United States from 2006 to 2013: social contagion or random clusters?},
  author={Lankford, Adam and Tomek, Sara},
  journal={Suicide and Life-threatening Behavior},
  volume={48},
  number={4},
  pages={459--467},
  year={2018},
  publisher={Wiley Online Library}
}

@article{towers2018detecting,
  title={Detecting the contagion effect in mass killings; a constructive example of the statistical advantages of unbinned likelihood methods},
  author={Towers, Sherry and Mubayi, Anuj and Castillo-Chavez, Carlos},
  journal={Plos one},
  volume={13},
  number={5},
  pages={e0196863},
  year={2018},
  publisher={Public Library of Science San Francisco, CA USA}
}
\end{document}